\documentclass[12pt]{article}
\usepackage{amsmath,amssymb}
\newcommand{\be}{\begin{equation}}
\newcommand{\ee}{\end{equation}}
\newcommand{\bea}{\begin{eqnarray}}
\newcommand{\eea}{\end{eqnarray}}

\newcommand{\p}[1]{(\ref{#1})}
\newcommand{\lb}{\label}

\begin{document}
\begin{titlepage}

\begin{center}
{\Large\bf Higher Spins in Harmonic Superspace} \vspace{1.5cm}

{\large\bf
Evgeny Ivanov}
\vspace{1cm}

{\it Bogoliubov Laboratory of Theoretical Physics,
JINR, \\
141980, Dubna, Moscow Region, Russia},\\
{\it Moscow Institute of Physics and Technology,\\
141700, Dolgoprudny, Moscow Region, Russia},\\
{\tt eivanov@theor.jinr.ru}\\[8pt]

\end{center}
\vspace{2cm}

\begin{abstract}
\noindent We report on a recent progress in constructing off-shell ${\cal N}=2, 4D$ supersymmetric integer
higher-spin theory in terms of unconstrained harmonic
analytic gauge superfields and their cubic interaction with the matter hypermultiplets. For even
superspins a new equivalent representation of the hypermultiplet couplings in terms of analytic $\omega$ superfield
is presented. It involves both cubic and quartic vertices.
\end{abstract}
\vspace{6cm}

\begin{center}
{\it Invited talk at the ``Academician A.A. Slavnov Memorial Conference'', \\
Moscow, December 21 - 22, 2022}
\end{center}

\end{titlepage}

\setcounter{page}{1}
\section{Introduction}
Supersymmetric higher-spin theories attract a lot of attention for many years (see, e.g., \cite{Vas1,Vas2,BekBul,Sagn,Did,Snow} and refs. therein). One of the basic reasons
of such an interest is that they can be considered as  a bridge between superstring theory and low-energy (super)gauge theories.

Free massless bosonic and fermionic higher spin field theories have
been constructed in \cite{Fron,FanFron}. The component description of {$4D$, ${\cal N}=1$} supersymmetric
free massless higher spin models was started in \cite{Courtright1979,Vasiliev1980}.
The natural tools of dealing with supersymmetric theories are off-shell superfield approaches.
When formulated in terms of superfields,  the supersymmetry is closed on the off-shell supermultiplets
with the correct sets of the auxiliary fields. Formulations in terms of unconstrained superfields are most preferable and suggestive,
especially in the quantum domain. The off-shell Lagrangian description of {$4D$}
free higher spin {${\cal N}=1$} models (both on the flat and on the AdS backgrounds)  through {${\cal N}=1$} superfields
has been given many years ago in refs. \cite{Kuzenko1, Kuzenko2} \footnote{See also \cite{KK1,KK2} where, in particular,
the relevant component off- and on-shell Lagrangians are discussed in more detail.}.

For long time, off-shell superfield Lagrangian formulations of higher-spin theories with extended
supersymmetry, such that all supersymmetries are manifest,  were not known even at the free level.
This problem  was recently solved in \cite{BIZ1}, where an off-shell manifestly {${\cal N}=2$}
supersymmetric unconstrained formulation of the Fronsdal theory for $4D$ integer spins has been constructed for the first time,
based upon the harmonic superspace approach \cite{HSS1,HSS}. Later on,  the manifestly {${\cal N}=2$} supersymmetric off-shell cubic couplings of {$4D, {\cal N}=2$}
higher-spin gauge superfields to the hypermultiplets were presented \cite{BIZ2} \footnote{See \cite{BIZZ} for a brief review.}. Recently, the more detailed analysis
of this theory, including its component structure, was undertaken in  \cite{BIZ3}.

Our papers \cite{BIZ1,BIZ2,BIZ3} launched a new promising direction of applications of the harmonic superspace formalism, that time
to higher-spin theories. In this report we briefly outline the basic points of this approach. The novel material is the reformulation
of the hypermultiplet couplings to higher spin ${\cal N}=2$ multiplets for even values of the superspin in terms of the so called
``$\omega$-representation'' for the analytic hypermultiplet superfields (section 4).


\section{Harmonic superspace}

Nowadays, the self-consistent off-shell superfield formalism for {${\cal N}=2$}
and {${\cal N}=3$} theories in four dimensions is the harmonic superspace approach \cite{HSS}.

The harmonic {${\cal N}=2$} superspace (HSS) is parametrized by the following coordinate set:
\bea
Z = (x^m\,,\;\theta^\alpha_i\,, \;\bar\theta^{\dot\alpha\,j}, u^{\pm i}), \quad u^{\pm i} \in SU(2)/U(1),  \; u^{+ i}u^-_i = 1. \lb{HSS}
\eea
Its fundamental feature is the presence of the analytic harmonic {${\cal N}=2$} superspace as an invariant subspace:
\bea
\zeta_A = (x^m_A, \theta^{+ \alpha}, \bar\theta^{+ \dot\alpha}, u^{\pm i}), \; \theta^{+ \alpha, \dot\alpha} := \theta^{\alpha, \dot\alpha i} u^+_i,
\; x^m_A := x^m - 2i\theta^{(i}\sigma^m \bar\theta^{j)}u^+_iu^+_j\,.\lb{AnHSS}
   \nonumber
\eea
The analytic basis of the full HSS \p{HSS} amounts to the coordinate set
\bea
Z = (\zeta_A, \theta^{- \alpha}, \bar\theta^{- \dot\alpha}), \quad \theta^{- \alpha, \dot\alpha} := \theta^{\alpha, \dot\alpha i} u^-_i\,. \lb{HSSan}
\eea

 All basic {${\cal N}=2$} off-shell supermultiplets are described by the analytic superfields:
\bea
\underline{\rm SYM}:&& V^{++}(\zeta_A)\,, \;\;  \underline{\rm matter \;hypermultiplets}: \; q^{+}(\zeta_A)\,, \,\bar{q}^{+}(\zeta_A) \nonumber \\
\underline{\rm supergravity}:&& h^{++ m}(\zeta_A)\,,\,h^{++ \alpha +}(\zeta_A)\,, \,h^{++ \dot\alpha +}(\zeta_A)\,, \,h^{++ 5}(\zeta_A)\,.  \nonumber
\eea

In what follows, we will need the expressions for various harmonic and spinor derivatives in the analytic basis:
\bea
&&{\cal  D}^{\pm\pm} = \partial^{\pm\pm} - 2i\theta^{\pm\alpha}\bar\theta^{\pm\dot\alpha}\partial_{\alpha\dot\alpha}
+ \theta^{\pm\alpha}\partial^\pm_\alpha + \bar\theta^{\pm\dot\alpha}\partial^\pm_{\dot\alpha} + i [(\theta^{\pm})^2  - (\bar\theta^{\pm})^2]\,\partial_5\,, \nonumber \\
&& {\cal D}^0 = \partial^{0} + \theta^{+\alpha}\partial^-_\alpha + \bar\theta^{+\dot\alpha}\partial^-_{\dot\alpha}
- \theta^{-\alpha}\partial^+_\alpha + \bar\theta^{-\dot\alpha}\partial^+_{\dot\alpha}\,, \nonumber \\
&& [{\cal D}^{++}, {\cal D}^{--}] = {\cal D}^0\,,\lb{HarmDer} \\
&& D^+_{\alpha} = \partial_\alpha^-\,, \quad \bar{D}^+_{\dot\alpha} = \partial_{\dot\alpha}^-\,, \label{D+} \\
&&   D^-_{\alpha} = [{\cal D}^{--}, D^+_{\alpha}] = - \partial_\alpha^- + 2i \bar{\theta}^{-\dot{\alpha}} \partial_{\alpha\dot{\alpha}} - 2i \theta^-_{\alpha} \partial_5\,, \nonumber \\
  &&  \bar{D}^-_{\dot{\alpha}} = [{\cal D}^{--}, \bar{D}^+_{\dot{\alpha}}] = -\partial_{\dot{\alpha}}^- - 2i \theta^{-\alpha} \partial_{\mu\dot{\alpha}}
  - 2i \bar{\theta}^-_{\dot{\mu}} \partial_5\,,\label{D-}
\end{eqnarray}
where
\bea
\partial^{\pm \pm} := u^{\pm i}\frac{\partial}{\partial u^{\mp i}}\,, \; \partial^{0} := u^{+i}\frac{\partial}{\partial u^{+ i}} - u^{-i}\frac{\partial}{\partial u^{- i}}\, \quad
\partial^\pm_\alpha := \frac{\partial}{\partial \theta^{\mp\alpha}}\,, \; \partial^\pm_{\dot\alpha} := \frac{\partial}{\partial \bar\theta^{\mp\dot\alpha}}\,. \nonumber
\eea
We included the partial derivatives with respect to an extra coordinate $x^5$ in these expressions,  which is necessary for describing the linearized ${\cal N}=2$ higher-spin gauge theories and massive hypermultiplets. In what follows,
neither ${\cal N}=2$ higher-spin gauge superfields nor the relevant gauge superparameters depend on this coordinate. With respect to ${\cal N}=2$ supersymmetry the coordinate $x^5$ in the analytic basis is assumed to transform as
\bea
\delta_\epsilon x^5 = 2i\big(\epsilon^-\theta^+ -
\bar\epsilon^-\bar\theta^+ \big), \lb{Tranfifth}
\eea
{\it i.e.}, the corresponding extension of the analytic subspace \p{AnHSS}, that is $(\zeta_A, x^5)$, stays closed under ${\cal N}=2$ supersymmetry.  The shortness of the covariant derivatives $D^+_{\alpha}, \bar{D}^+_{\dot\alpha}$
in the analytic basis reflects the independence of the analytic superfields  of the coordinates $\theta^{-\alpha}, \bar\theta^{-\dot\alpha}$ in this basis. Also, these spinor derivatives commute
with the harmonic derivative ${\cal D}^{++}$, which implies that the latter preserves ${\cal N}=2$ harmonic analyticity.

\subsection{${\cal N}=2$ spin 1}

The simplest example is supplied by Abelian {${\cal N}=2$} gauge theory, with the analytic gauge potential $V^{++}(\zeta_A)$ as the basic entity,
\bea
V^{++}(\zeta_A)\,, \quad \delta V^{++} = {\cal D}^{++}\Lambda (\zeta_A)\,. \;
\eea
In Wess-Zumino (WZ) gauge,
\bea
&& V^{++}(\zeta_A) = (\theta^+)^2 \phi + (\bar\theta^+)^2 \bar\phi + 2i\theta^{+\alpha}\bar\theta^{+\dot\alpha} A_{\alpha\dot\alpha} \nonumber \\
&& +\, (\bar\theta^+)^2 \theta^{+\alpha}\psi_\alpha^i u^-_i +
(\theta^+)^2 \bar\theta^{+}_{\dot\alpha}\bar\psi^{\dot\alpha i} u^-_i +(\theta^+)^2(\bar\theta^+)^2 D^{(ik)}u^-_iu^-_k \,,\nonumber
\eea
{$4D$} fields {$\phi,\, \bar\phi\,, \,A_{\alpha\dot\alpha}\,, \,\psi_\alpha^i\,, \, \bar\psi_{\dot\alpha}^i \,, \,D^{(ik)}$}
constitute an abelian gauge {${\cal N}=2$} off-shell multiplet ({ 8 + 8} off-shell degrees of freedom).

The invariant action is given by the expression:
\bea
S \sim \int d^{12}Z\,\, \big(V^{++} V^{--}\big)\,, \; {\cal D}^{++} V^{--} - {\cal D}^{--} V^{++} = 0\,, \; \delta V^{--} = D^{--}\Lambda\,, \lb{Spin1Act}
\eea
where
\bea
d^{12}Z := [du]  d^4x d^4 \theta^+ d^4 \theta^- \nonumber
\eea
is the measure of integration over the full HSS, with $[du]$ denoting integration over harmonics. The second (non-analytic) harmonic connection $V^{--}$
is specified, up to the analytic parameter gauge freedom, in terms of $V^{++}(\zeta_A)$  by the harmonic ``flatness'' condition
in \p{Spin1Act}\footnote{The expression for $V^{--}$ can be obtained
either by solving the flatness condition with the help of harmonic distributions (see book \cite{HSS}) or by a direct calculation in the component formalism,
starting from the WZ form of $V^{++}$.}.

\subsection{${\cal N}=2$ spin 2: linearized ${\cal N}=2$ supergravity}

In this case, the analogs of {$V^{++}(\zeta_A)$} are the following set of analytic gauge potentials:
\bea
&& \Big( h^{++m}(\zeta_A)\,, \;h^{++5}(\zeta_A)\,, \; h^{++\hat{\mu}+}(\zeta_A) \Big),  \quad \hat{\mu} = (\mu\,, \dot{\mu})\,, \nonumber \\
&&\delta_\lambda h^{++m } = {\cal D}^{++} \lambda^m + 2i \big( \lambda^{+\alpha} \sigma^m_{\alpha\dot{\alpha}} \bar{\theta}^{+\dot{\alpha}}
        + \theta^{+\alpha} \sigma^m_{\alpha\dot{\alpha}} \bar{\lambda}^{+\dot{\alpha}}\big)\,, \nonumber \\
&&\delta_\lambda h^{++5} = {\cal D}^{++} \lambda^5 - 2i \big(\lambda^{+{\alpha}} \theta^{+}_{\alpha} - \bar\theta^{+}_{\dot{\alpha}}\bar\lambda^{+\dot{\alpha}}\big),
\delta_\lambda h^{++\hat{\mu}+} = {\cal D}^{++} \lambda^{+\hat{\mu}}\,.  \lb{N2SG}
\eea

The WZ gauge is attained as
\begin{eqnarray}
&&h^{++m}
       =
        -2i \theta^+\sigma^a \bar{\theta}^+ \Phi^m_a
       +  \big[(\bar{\theta}^+)^2 \theta^+ \psi^{m\,i}u^-_i + c.c.\big]+ \ldots  \nonumber \\
&&h^{++5} =
       -2 i \theta^+ \sigma^a \bar{\theta}^+ C_a + \ldots\,, \quad h^{++\mu+} = \ldots\,,  \nonumber
\end{eqnarray}
where $\dots $  stand for auxiliary fields\footnote{The complete form of WZ gauge can be found in ref. \cite{BIZ1}.}. The residual gauge freedom in WZ gauge is constituted by\footnote{The spinor indices are raised
and lowered in the standard way, with the help of antisymmetric tensor $\varepsilon_{\mu\nu}, \varepsilon_{12} = -\varepsilon^{12} = 1$, e.g., $l_{(\nu}^{\;\;\;\mu)} = \varepsilon_{\nu\rho}l^{(\rho\mu)}$.}:
\begin{eqnarray}
\lambda^m \;\Rightarrow\; a^m(x)\,, \; \lambda^5 \, \Rightarrow \; b(x)\,, \;
\lambda^{\mu+} \;\Rightarrow \; \epsilon^{\mu i}(x) u^+_i + \theta^{+\nu}l_{(\nu}^{\;\;\;\mu)}(x)\,.\nonumber
\end{eqnarray}

One ends with the standard field content $40 + 40$ of minimal ${\cal N}=2$ Einstein  supergravity \cite{MinN2SG,nider,nider2}.
The physical fields are $\Phi^m_a, \psi^{m\,i}_\mu, C_a$ (spins ${\bf (2, 3/2, 3/2, 1)}$ on shell).
The spin 1 part of {$\Phi^m_a$}
is gauged away by the local ``Lorentz'' parameters {$l_{(\nu}^{\;\;\;\mu)}(x),  l_{(\dot{\nu}}^{\;\;\;\dot{\mu})}(x)$}.
In this ``physical'' gauge
\bea
&& \Phi^m_a \sim \Phi_{\beta\dot\beta\alpha\dot\alpha} \Rightarrow \Phi_{(\beta\alpha)(\dot\beta\dot\alpha)}
+ \varepsilon_{\alpha\beta}\varepsilon_{\dot\alpha\dot\beta} \Phi\,.\nonumber
\eea
The invariant action reads
\bea
&& S_{(s=2)} = -\int d^4x d^8\theta du \Big( G^{++\alpha\dot\alpha}G^{--}_{\alpha\dot\alpha} + G^{++5}G^{--5}\Big), \lb{LinN2Act} \\
&&G^{++ \mu\dot\mu} := h^{++ \mu\dot\mu} + 2i\big( h^{++\mu+}\bar\theta^{-\dot\mu} + \theta^{-\mu}h^{++\dot\mu+}\big)\,, \nonumber \\
&&G^{++ 5} := h^{++ 5} - 2i\big( h^{++\mu+}\theta^-_\mu - \bar\theta^-_{\dot\mu}h^{++\dot\mu+}\big)\,,\nonumber \\
&& {\cal D}^{++}G^{--\mu\dot\mu} = {\cal D}^{--}G^{++\mu\dot\mu}\,, \quad {\cal D}^{++}G^{--5} = {\cal D}^{--}G^{++5}\,. \label{Spin2Zero}
\eea
Note that the flatness conditions \p{Spin2Zero} (likewise the analogous conditions for higher ${\cal N}=2$ spins ${\bf s}$)  amount to those involving analytic
potentials. These latter equations are uniquely solved as in the case of the spin ${\bf s}=1$.

After passing to components, the spin 2 part of the Lagrangian can be written as:
\bea
&&G^{++\alpha\dot\alpha}_{(\Phi)}G_{(\Phi)\alpha\dot\alpha}^{--} +  G_{(\Phi)}^{++5}G_{(\Phi)}^{--5} \quad \Rightarrow \nonumber\\
&& {\cal L}_{(\Phi 2)} =
-\frac{1}{4}\Big[\Phi^{(\alpha\beta)(\dot{\alpha}\dot{\beta})} \Box
\Phi_{(\alpha\beta)(\dot{\alpha}\dot{\beta})} -
\Phi^{(\alpha\beta)(\dot{\alpha}\dot{\beta})}
\partial_{\alpha\dot{\alpha}} \partial^{\rho\dot{\rho}}
\Phi_{(\rho\beta)(\dot{\rho}\dot{\beta})} \nonumber\\
 &&
\;\;\;\;\;\;\;\;\;\;\;         + 2\, \Phi
\partial^{\alpha\dot{\alpha}} \partial^{\beta\dot{\beta}}
\Phi_{(\alpha\beta)(\dot{\alpha}\dot{\beta})}
        - 6 \Phi \Box \Phi \Big].\nonumber
\eea
It is invariant under the linearized gauge spin 2 transformations $\delta \Phi_{\beta \dot{\beta}\alpha \dot{\alpha}} =\frac{1}{2} \left(\partial_{\alpha\dot{\alpha}} a_{\beta\dot{\beta}}
+ \partial_{\beta\dot{\beta}} a_{\alpha\dot{\alpha}}  \right),$  $\delta \Phi = \frac{1}{4} \partial_{\alpha\dot{\alpha}} a^{\alpha\dot{\alpha}}\,.$

\setcounter{equation}{0}

\section{${\cal N}=2$ spin 3 and higher spins}

A generalization to higher integer spin {${\cal N}=2$} supermultiplets is rather straightforward. The spin 3 example is instructive.

The spin 3 triplet  of analytic gauge superfields is introduced as
\bea
h^{++(\alpha\beta)(\dot\alpha\dot\beta)} (\zeta)\,, \; h^{++ \alpha\dot\alpha}(\zeta), \; h^{++(\alpha\beta)\dot{\alpha}+}(\zeta), \;
h^{++(\dot\alpha\dot\beta){\alpha}+}(\zeta)\,, \nonumber
\eea
and has the following transformation laws, with the analytic gauge parameters,
\bea
&& \delta h^{++(\alpha\beta)(\dot{\alpha}\dot{\beta})} = {\cal D}^{++} \lambda^{(\alpha\beta)(\dot{\alpha}\dot{\beta})}
       + 2i \big[\lambda^{+(\alpha\beta)(\dot{\alpha}}  \bar{\theta}^{+\dot{\beta})}
        + \theta^{+(\alpha}  \bar{\lambda}^{+\beta)(\dot{\alpha}\dot{\beta})}\big],
        \nonumber \\
       && \delta h^{++\alpha\dot{\alpha}} = {\cal D}^{++} \lambda^{\alpha\dot{\alpha}} - 2i  \big[\lambda^{+(\alpha\beta)\dot{\alpha}} \theta^{+}_{\beta} +
        \bar\lambda^{+(\dot\alpha\dot\beta){\alpha}} \bar\theta^{+}_{\dot\beta}\big], \nonumber \\
        && \delta h^{++(\alpha\beta)\dot{\alpha}+} = {\cal D}^{++} \lambda^{+(\alpha\beta)\dot{\alpha}}\,, \; \delta h^{++(\dot{\alpha}\dot{\beta})\alpha+}
        = {\cal D}^{++} \lambda^{+(\dot{\alpha}\dot{\beta})\alpha}\,. \lb{N2spin3}
        \eea

The bosonic physical fields in  the WZ gauge are collected in the $\theta^+, \bar\theta^+$ expansions
\bea
h^{++(\alpha\beta)(\dot{\alpha}\dot{\beta})}
        &=&
        -2i \theta^{+\rho} \bar{\theta}^{+\dot{\rho}} \Phi^{(\alpha\beta)(\dot{\alpha}\dot{\beta})}_{\rho\dot{\rho}} + [(\bar\theta^+)^2 \theta^+
        \psi^{(\alpha\beta)(\dot\alpha\dot\beta)i} u^-_i + c.c.] + \ldots \nonumber \\
h^{++\alpha\dot{\alpha}} &=&
        -2i \theta^{+\rho} \bar{\theta}^{+\dot{\rho}} C^{\alpha\dot{\alpha}}_{\rho\dot{\rho}} + \ldots \nonumber
\eea
The physical gauge fields are {$\Phi^{(\alpha\beta)(\dot{\alpha}\dot{\beta})}_{\rho\dot{\rho}}$} (spin 3 gauge field),
{$C^{\alpha\dot{\alpha}}_{\rho\dot{\rho}}$} (spin 2 gauge field) and {$\psi^{(\alpha\beta)(\dot{\alpha}\dot{\beta})i}_\gamma$}
(spin 5/2 gauge field). The rest of fields are auxiliary (like in other cases, the full form of WZ gauge for ${\bf s}=3$ can be found in \cite{BIZ1}).
On shell, one ends up with the multiplet
{$({\bf 3, 5/2, 5/2, 2})$}.

Some residual gauge freedom can be used to reach the following irreducible form for physical bosonic gauge fields
\bea
&& \Phi_{\gamma\dot\gamma (\alpha\beta)(\dot{\alpha}\dot{\beta})} = \Phi_{(\alpha\beta\gamma)(\dot\alpha\dot\beta\dot\gamma)}
+ \varepsilon_{\dot\gamma(\dot\alpha} \varepsilon_{\gamma(\beta} \Phi_{\alpha)\dot\beta)}\,, \nonumber \\
&&  C_{\gamma\dot\gamma \alpha\dot\alpha} = C_{(\gamma\alpha)(\dot\gamma\dot\alpha)} + \varepsilon_{\gamma\alpha}\varepsilon_{\dot\gamma\dot\alpha} C\,, \nonumber
\eea
with the residual gauge transformations
\bea
&&\delta \Phi_{(\alpha \gamma \beta ) (\dot{\alpha} \dot{\gamma} \dot{\beta} )}
    =
    \partial_{(\beta(\dot{\beta}} a_{\alpha \gamma) \dot{\alpha} \dot{\gamma})}\,, \quad \delta \Phi_{\alpha\dot{\beta}}
   = \frac{4}{9} \partial^{\gamma \dot{\gamma}} a_{(\alpha\gamma)(\dot{\beta}\dot{\gamma})}\,, \nonumber \\
&&  \delta C_{(\alpha\beta) (\dot{\alpha}\dot{\beta})} = \partial_{(\beta (\dot{\beta}} b_{\alpha)\dot{\alpha})},
    \quad
    \delta C = \frac{1}{4} \partial_{\alpha\dot{\alpha}} b^{\alpha\dot{\alpha}}. \nonumber
\eea
These are the correct gauge transformations of the Fronsdal spin 3 fields ({$\Phi_{(\alpha\beta\gamma)(\dot\alpha\dot\beta\dot\gamma)}, \Phi_{\alpha\dot\beta}$})
and spin 2 fields ({$C_{(\alpha\beta) (\dot{\alpha}\dot{\beta})}, C $}).

The invariant superfield action is constructed on the pattern of the spin 2 case
\bea
S_{(s=3)} = \int d^4x d^8\theta du \,\Big\{G^{++(\alpha\beta)(\dot\alpha\dot\beta)}G^{--}_{(\alpha\beta)(\dot\alpha\dot\beta)}
+ G^{++\alpha\dot\beta}G^{--}_{\alpha\dot\beta} \Big\}, \lb{Spin3Act}
\eea
where
\bea
&& G^{++(\alpha\beta)(\dot\alpha\dot\beta)} = h^{++(\alpha\beta)(\dot\alpha\dot\beta)}+ 2i\big[ h^{++(\alpha\beta)(\dot\alpha+} \bar\theta^{-\dot\beta)} -
h^{++(\dot\alpha\dot\beta) (\alpha +}\theta^{-\beta)}\big]\,, \nonumber\\
&&G^{++\alpha\dot\beta} = h^{++\alpha\dot\beta} - 2i \big[h^{++(\alpha\beta)\dot\beta+}\theta^-_\beta - \bar\theta^-_{\dot\alpha}h^{++(\dot\alpha\dot\beta)\alpha+} \big],
\nonumber \\
&& {\cal D}^{++} G^{--(\alpha\beta)(\dot\alpha\dot\beta)} - {\cal D}^{--}G^{++(\alpha\beta)(\dot\alpha\dot\beta)} = 0\,, \; {\cal D}^{++}G^{--\alpha\dot\beta} - {\cal D}^{--}G^{++\alpha\dot\beta} = 0. \lb{ZsroSpin3}
\eea

The component spin 3 bosonic action can be easily extracted and it reads
\bea
        && S_{(\Phi 3)} =   \int d^4x \Big\{\Phi^{(\alpha_1\alpha_2\alpha_3)( \dot{\alpha}_1\dot{\alpha}_2\dot{\alpha}_3)}
 \Box \Phi_{(\alpha_1\alpha_2\alpha_3)( \dot{\alpha}_1\dot{\alpha}_2\dot{\alpha}_3)} \nonumber \\
 && \;\;\;\;\;\;\;\;\;- \,\frac{3}{2} \Phi^{(\alpha_1\alpha_2\alpha_3)( \dot{\alpha}_1\dot{\alpha}_2\dot{\alpha}_3)}
            \partial_{\alpha_1\dot{\alpha}_1} \partial^{\rho\dot{\rho}} \Phi_{(\rho\alpha_2\alpha_3)( \dot{\rho}\dot{\alpha}_2\dot{\alpha}_3)} \nonumber \\
 &&\;\;\;\;\;\;\;\;\;+ \,3 \Phi^{(\alpha_1\alpha_2\alpha_3)( \dot{\alpha}_1\dot{\alpha}_2\dot{\alpha}_3)} \partial_{\alpha_1\dot{\alpha}_1} \partial_{\alpha_2\dot{\alpha}_2}
 \Phi_{\alpha_3\dot{\alpha}_3}
            - \frac{15}{4} \Phi^{\alpha\dot{\alpha}} \Box  \Phi_{\alpha\dot{\alpha}} \nonumber \\
           && \;\;\;\;\;\;\;\;\;+\,
            \frac{3}{8} \partial_{\alpha_1\dot{\alpha}_1} \Phi^{\alpha_1\dot{\alpha}_1} \partial_{\alpha_2\dot{\alpha}_2} \Phi^{\alpha_2\dot{\alpha}_2}\Big\}. \nonumber
        \eea

The general case with the maximal spin ${\bf s}$ is governed by the following analytic gauge potentials
\bea
h^{++\alpha(s-1)\dot\alpha(s-1)}(\zeta), h^{++\alpha(s-2)\dot\alpha(s-2)}(\zeta), h^{++\alpha(s-1)\dot\alpha(s-2)+}(\zeta),
h^{++\dot\alpha(s-1)\alpha(s-2)+}(\zeta), \nonumber
\eea
where {$\alpha(s) := (\alpha_1 \ldots \alpha_s), \dot\alpha(s) := (\dot\alpha_1 \ldots \dot\alpha_s)$}.
The relevant gauge transformations can also be defined and shown to leave, in the WZ gauge, the  physical field multiplet
{$({\bf s, s-1/2, s-1/2, s-1})$}.

 The invariant action has the universal form for any ${\bf s}$
\bea
&& S_{(s)} = (-1)^{s+1} \int d^4x
d^8\theta du \,\Big\{G^{++\alpha(s-1)\dot\alpha(s-1)}G^{--}_{\alpha(s-1)\dot\alpha(s-1)} \nonumber \\
&&\;\;\;\;\;  +\,
G^{++\alpha(s-2)\dot\alpha(s-2)}G^{--}_{\alpha(s-2)\dot\alpha(s-2)}
\Big\},  \lb{SpinsAct}
\eea
where
\bea
&& G^{\pm\pm\alpha(s-1)\dot\alpha(s-1)} = h^{\pm\pm\alpha(s-1)\dot\alpha(s-1)} + 2i \big[h^{\pm\pm\alpha(s-1)(\dot\alpha(s-2)+}\bar\theta^{-\dot\alpha_{s-1})} \nonumber \\
&& \;\;\;\;\;\;\;\;\;\;\;\;\;\;\;\;\;\;\;\;\;\;\;\;\;\;- h^{\pm\pm\dot\alpha(s-1)(\alpha(s-2)+}\,\theta^{-\alpha_{s-1})} \big], \nonumber \\
&& G^{\pm\pm\alpha(s-2)\dot\alpha(s-2)} = h^{\pm\pm\alpha(s-2)\dot\alpha(s-2)} - 2i \big[h^{\pm\pm(\alpha(s-2)\alpha_{s-1}) \dot\alpha(s-2)+}\theta^{-}_{\alpha_{s-1}} \nonumber \\
&& \;\;\;\;\;\;\;\;\;\;\;\;\;\;\;\;\;\;\;\;\;\;\;\;\;\;+ h^{\pm\pm\alpha(s-2)(\dot\alpha(s-2)\dot\alpha_{(s-1)})+}\,\bar\theta^{-}_{\alpha_{s-1}} \big], \lb{Gdef}
\eea
and the negatively charged potentials are related to the basic ones by the appropriate harmonic zero-curvature conditions.

\setcounter{equation}{0}

\section{Interactions with hypermultiplet}

 The construction of interactions of
higher spins is a very important (albeit difficult) task. The simplest interaction is described by a cubic vertex which is bilinear in the matter fields and of the first order
in gauge fields. At present, there is a vast literature on cubic higher spin interactions
(see, e.g., \cite{Bengtsson1983,FrMet1991,Met1993,ManvMkrRuehl,Fotopoulos:2007yq,Bekaert:2009ud,Khabarov:2020bgr,Khabarov:2020deh}).

 Supersymmetric {$\mathcal{N}=1$} extensions of the purely bosonic
cubic vertices with matter and the corresponding supercurrents were
constructed in terms of {$\mathcal{N}=1$} superfields, e.g., in refs. \cite{GKKBB}.

In ref. \cite{BIZ2} there were constructed, for the first time, off-shell
manifestly {$\mathcal{N}=2$} supersymmetric cubic couplings
$\mathbf{\frac{1}{2}, \frac{1}{2}, s}$  of an arbitrary higher
integer  superspin {$\mathbf{s}$} gauge {$\mathcal{N}=2$} multiplet to the
hypermultiplet matter (superspin $\mathbf{\frac{1}{2}}$ ) in {$4D, \mathcal{N}=2$} harmonic
superspace.

The starting point was the {${\cal N}=2$} hypermultiplet off-shell free action:
\begin{equation}
    S = \int  d\zeta^{(-4)}  \; \mathcal{L}^{+4}_{free} = -\int d\zeta^{(-4)}  \; \frac{1}{2} q^{+a} \mathcal{D}^{++} q^+_a, \; a = 1,2, \lb{Freeq}
\end{equation}
where $d\zeta^{(-4)} = [du] d^4x_A d^4\theta^+$ is the measure of integration over the analytic superspace.
The doublet indices $a$ are rotated by an extra (``Pauli-G\"ursay'') group ${\rm SU}(2)_{PG}$ commuting with ${\cal N}=2$ supersymmetry and the automorphism ${\rm SU}(2)$.

The hypermultiplet described by the action \eqref{Freeq} can be either massive or massless.
In the case of massive $\mathcal{N}=2$ hypermultiplet (without gauge fields) the minimal mechanism of generating
mass is through a non-zero central charge in ${\cal N}=2$ superalgebra, with mass being the eigenvalue
of the central charge generator.
The common way to introduce the central charge is to add an extra fictitious coordinate $x^5$,
as was already discussed in Sect.2. The hypermultiplet is assumed to depend on $x^5$ almost trivially:
\begin{equation}\label{global-1}
  q^{+a}(\zeta_A, x^5) = (e^{-im \tau^3 x^5})^a_b\,  q^{+ b}(\zeta_A, u) \quad \Leftrightarrow  \quad  \partial_5 q^{+a} := -im (\tau^3)^a_{\;b} q^{+b}\,,
\end{equation}
where
\begin{equation}\label{tau3}
    (\tau_3)^a_{\;b}
    =
    {\tiny  \begin{pmatrix}
            1 & 0 \\
            0 & -1
    \end{pmatrix}}
    ,
    \;\;\;\;\;\;\;\;\;
    (\tau_3)_{ab} = \epsilon_{ac}   (\tau_3)^c_{\;b}
    =
    {\tiny  \begin{pmatrix}
            0 & -1 \\
            -1 & 0
    \end{pmatrix}}.
\end{equation}
For avoiding any integration over $x^5$ in the hypermultiplet action \eqref{Freeq},
we impose the standard Scherk-Schwarz condition that
$\partial_5 q^{+a}$ coincides with the action
of some ${\rm U}(1)_{PG} \subset {\rm SU}(2)_{PG}$. The parameter $m$ is a mass of the hypermultiplet,
so we deal with the massive hypermultiplet in the general case. It is evident that
\be
\partial_5 (q^{+a}\mathcal{D}^{++} q^+_a) = 0\,. \label{x5independence}
\ee
The action \eqref{Freeq} is invariant under the rigid $\mathcal{N}=2$ supersymmetry,
\begin{equation}\label{sup-hyp}
    \delta^*_{\epsilon} q^{+a} = -\delta_\epsilon Z^M\partial_M q^{+a}\,,
\end{equation}
where we need to take into account also the transformation of $x^5$ \eqref{Tranfifth} for $m\neq 0$. Because of the property
\eqref{x5independence}, introducing a mass does not affect $\mathcal{N}=2$ supersymmetry
transformation of the Lagrangian, though modifies  $\mathcal{N}=2$ transformation of $q^{+a}$. The internal symmetry
of the single hypermultiplet action is ${\rm SU}(2)_{PG} \times{\rm SU}(2)_{aut}$ in the massless case
($\partial_5 q^{+a} =0$) and ${\rm U}(1)_{PG} \times{\rm SU}(2)_{aut}$ in the massive case ($\partial_5 q^{+a} \neq 0$).

We shall reproduce the gauge higher-spin {${\cal N}=2$} superfields from gauging the appropriate higher-derivative rigid (super)symmetries of the free
hypermultiplet action \eqref{Freeq}. In this section, for simplicity, we put all coupling constants equal to unity.\\

\noindent$\mathbf{Superspin\,\, 1}$. The simplest symmetry is of zero order in derivatives, it is {${\rm U}(1)$} transformation of {$q^{+a}$},
\begin{equation}
\delta q^{+a} = -\lambda_0 J q^{+ a}, \quad J q^{+ a} = i (\tau_3)^a_{\;b} q^{+b}\,.\lb{U1}
\end{equation}

Gauging of this symmetry amounts just to replacing {$\lambda_0$} by an analytic superparameter, {$\lambda_0 \rightarrow  \lambda(\zeta)$} and, in order to make the
{$q^{+a}$} action gauge-invariant, to performing the appropriate replacement in \eqref{Freeq}:
\bea
&& {\cal D}^{++} \;\Rightarrow \;\mathfrak{D}^{++}_{(1)} =  {\cal D}^{++} + \hat{\cal H}^{++}_{(1)}\,, \quad  \hat{\cal H}^{++}_{(1)}= h^{++}J\,, \label{Spin1coupl} \\
&& \delta_\lambda \hat{\cal H}^{++}_{(1)} = [{\cal D}^{++}, \hat{\Lambda}_{(1)}]\,, \quad \hat{\Lambda}_{(1)}= \lambda J\; \Rightarrow \;
\delta_\lambda h^{++} = {\cal D}^{++}\lambda\,. \nonumber
\eea
There is no direct relation between {$J$} and {$\partial_5$}: one can choose either {$\partial_5 q^{+a}=0$}, that corresponds to
 massless {$q^{+a}$}
or {$\partial_5 q^{+a} \sim   mJ q^{+ a}$}, that
corresponds to massive {$q^{+a}$}.\\

\noindent$\mathbf{Superspin\,\, 2}$. The rigid symmetry to be gauged in the spin 2 case is more complicated, it is of the first order in derivatives
\begin{equation}
    \delta_{rig} q^{+a} =  -\hat{\Lambda}_{rig} q^{+a},\nonumber
\end{equation}
\begin{eqnarray}
    \hat{\Lambda}_{rig} &=& \left( \lambda^{\alpha\dot{\alpha}}_0 - 2i \lambda^{-\alpha}_0 \bar{\theta}^{+\dot{\alpha}} - 2i \theta^{+\alpha} \bar{\lambda}^{-\dot{\alpha}}_0  \right) \partial_{\alpha\dot{\alpha}}
    + \lambda^{+\alpha}_0 \partial^-_{\alpha} + \bar{\lambda}^{+\dot{\alpha}}_0 \partial^-_{\dot{\alpha}} \nonumber \\
   &&+\, \left( \lambda^5_0 + 2i \lambda^{\hat{\alpha}-}_0 \theta^+_{\hat{\alpha}} \right) \partial_5  := \Lambda^M_0 \partial_M\,, \quad [{\cal D}^{++},\hat{\Lambda}_{rig}] = 0\,.\nonumber
\end{eqnarray}
It is spanned by five constant bosonic parameters {$\lambda^{\alpha\dot{\alpha}}_0$}, {$\lambda^5_0$}, four constant spinor parameters
{$\lambda^{\pm\hat{\alpha}_0} = \lambda^{\hat{\alpha}i}_0 u^\pm_i$}
and can be treated as a copy of the rigid {${\cal N}=2$} supersymmetry transformations in their active form.
We gauge just these transformations and leave intact the rigid {${\cal N}=2$} supersymmetry realized on the superspace coordinates.

In the spin 2 case there are two options for gauge transformations of the hypermultiplet:
\begin{equation}
    \delta_1 q^{+a} =  -\hat{\Lambda}_{(2)} q^{+a},
    \;\;\;\;\;
    \hat{\Lambda}_{(2)}: = \lambda^M \partial_M = \lambda^{\alpha\dot{\alpha}} \partial_{\alpha\dot{\alpha}}
    + \lambda^{+\alpha} \partial^-_{\alpha} + \bar{\lambda}^{+\dot{\alpha}} \partial^-_{\dot{\alpha}} + \lambda^5 \partial_5\,,\lb{Spiin2first}
\end{equation}
\begin{equation}
    \delta_2 q^{+a} =  -\frac{1}{2} \Omega_{(2)} q^{+a},
    \;\;\;\;\;
    \Omega_{(2)} := (-1)^{P(M)} \partial_M \lambda^M = \partial_{\alpha\dot{\alpha}}\lambda^{\alpha\dot{\alpha}}
     - \partial^-_{\alpha} \lambda^{+\alpha}  - \partial^-_{\dot{\alpha}}\bar{\lambda}^{+\dot{\alpha}}\,,\lb{Spin2second}
\end{equation}
\begin{equation}
    \left(\delta_1 + \delta_2 \right) \mathcal{L}^{+4}_{free}
    =
     \frac{1}{2} q^{+a} [\mathcal{D}^{++}, \hat{\Lambda}_{(2)}] q^+_a. \nonumber
\end{equation}

As the next step, we introduce the differential operator
\begin{equation}
    \hat{\mathcal{H}}^{++}_{(2)} = h^{++\alpha\dot{\alpha}} \partial_{\alpha\dot{\alpha}} + h^{++\hat{\mu}+} \partial^-_{\hat{\mu}} + h^{++5}\partial_5\,,\lb{H++2}
\end{equation}
with the transformation law
\begin{equation}
    \delta \hat{\mathcal{H}}^{++}_{(2)} = [\mathcal{D}^{++}, \hat{\Lambda}_{(2)}].\lb{N2LinTrans}
\end{equation}
Then the linear in gauge superfields part of the gauge-invariant action is expressed as
\begin{equation}
    \mathcal{L}^{+4}_{free}\;\;\to\;\; \mathcal{L}^{+4 (s=2)}_{gauge} = \mathcal{L}^{+4}_{free} - \frac{1}{2} q^{+a} \hat{\mathcal{H}}^{++}_{(2)} q^+_a\,.\lb{Spin2Hyper}
\end{equation}

Note that {$\mathcal{L}^{+4 (s=2)}_{gauge}$} can be made fully gauge invariant (and not only at the linearized level) by modifying the gauge
transformation law \p{N2LinTrans} of $\hat{\mathcal{H}}^{++}_{(2)}$ in the following way
\begin{equation}
    \delta_{full} \hat{\mathcal{H}}^{++}_{(2)}  =
    [\mathcal{D}^{++} +\hat{\mathcal{H}}^{++}_{(2)}, \hat{\Lambda}_{(2)}]\,.\label{FullInv}
\end{equation}

The complete nonlinear harmonic superfield action of ${\cal N}=2$ supergravity including the pure supergravity nonlinear
part was constructed in \cite{GNKES}.\\

\noindent$\mathbf{Superspin\,\, 3}$. A surprising feature of the superspin 3 case is that the relevant rigid two-derivative transformations to be gauged and
the resulting couplings of the relevant gauge superfields to the hypermultiplet $q^{+ a}$ can be consistently defined only at cost of breaking rigid {${\rm SU}(2)_{PG}$} symmetry
down to {${\rm U}(1)$} generator which is manifestly present in all relations. This peculiar property extends to all odd {${\cal N}=2$} spins.

Though in the spin ${\bf 3}$ case  one can define from the very beginning, 4 independent gauge transformations of {$q^{+ a}$}, only two their fixed combinations can be compensated
by the appropriate transformations of gauge superfields:
\begin{eqnarray}
\delta_{\lambda} q^{+a} &=& -[ \lambda^{\alpha\dot\alpha M}\partial_M\partial_{\alpha\dot\alpha} + \frac12 ( \partial_{\alpha\dot\alpha}\lambda^{\alpha\dot\alpha M})\partial_M
+ \frac12\Omega^{\alpha\dot\alpha}_{(3)}\partial_{\alpha\dot\alpha} +\frac12\Omega_{(3)}] (\tau_3)^a_{\;b} q^{+ b}\nonumber\\
&& \delta_{\xi} q^{+a} =  - \xi \,\Omega_{(3)}\,J q^{+a} = - i\xi \,\Omega_{(3)}\,(\tau_3)^a_{\;b} q^{+ b}\,,\lb{Spin4GaugeqTran}
\end{eqnarray}
where\footnote{Hereafter, all Lorentz indices of the same nature in the coefficients of the differential operators are assumed to be properly symmetrized with those hidden in the multi-index $M$.}
\begin{eqnarray}
&& \lambda^{\alpha\dot\alpha M}\partial_M = \lambda^{(\alpha\beta)(\dot{\alpha}\dot{\beta})}\partial_{\beta\dot{\beta}} + \lambda^{(\alpha\beta)\dot{\alpha}+} \partial^-_\beta +
    \bar{\lambda}^{(\dot{\alpha}\dot{\beta})\alpha+} \partial^-_{\dot{\beta}} + \lambda^{\alpha\dot{\alpha}} \partial_5\,, \nonumber\\
&&\Omega_{(3)} = (\partial_{\alpha\dot\alpha}\Omega_{(3)}^{\alpha\dot\alpha})\,, \quad \Omega^{\alpha\dot\alpha}_{(3)} = (-1)^{P(M)}(\partial_M\lambda^{\alpha\dot\alpha M})\,. \nonumber
\end{eqnarray}

After defining
\begin{eqnarray}
&& \hat{\mathcal{H}}^{++\alpha\dot{\alpha}} = h^{++(\alpha\beta)(\dot{\alpha}\dot{\beta})}\partial_{\beta\dot{\beta}} + h^{++(\alpha\beta)\dot{\alpha}+} \partial^-_\beta
    +
    \bar{h}^{++(\dot{\alpha}\dot{\beta})\alpha+} \partial^-_{\dot{\beta}} + h^{++\alpha\dot{\alpha}} \partial_5\,, \nonumber \\
&& \Gamma^{++\alpha\dot{\alpha}} = \partial_{\beta\dot{\beta}}h^{++(\alpha\beta)(\dot{\alpha}\dot{\beta})}-
    \partial^-_{\beta}h^{++(\alpha\beta)\dot{\alpha}+} -
    \partial^-_{\dot{\beta}}
    h^{++\alpha(\dot{\alpha}\dot{\beta})+}, \nonumber \\
 &&  \hat{\mathcal{H}}^{++}_{(3)}
     = \hat{\mathcal{H}}^{++\alpha\dot{\alpha}} \partial_{\alpha\dot{\alpha}}, \quad \Gamma^{++}_{(3)} = \partial_{\alpha\dot{\alpha}}\Gamma^{++\alpha\dot{\alpha}}\,, \nonumber \\
&&\delta \hat{\mathcal{H}}^{++}_{(3)} = [\mathcal{D}^{++}, \hat{\Lambda}_{(3)}]\,, \,
\hat{\Lambda}_{(3)} = {\lambda}^{\alpha\dot{\alpha}M}\partial_M \partial_{\alpha\dot{\alpha}}\,, \;
\delta\Gamma^{++}_{(3)} = {\cal D}^{++}\Omega_{(3)}\,,  \label{Gaugespin3}
\end{eqnarray}
we can construct a gauge invariant extension of the {$q^+$} action as
\begin{eqnarray}
\mathcal{L}^{+4(s=3)}_{gauge} = \mathcal{L}^{+4}_{free}
-\frac12q^{+a} \left(\mathcal{D}^{++} + \hat{\mathcal{H}}^{++}_{(3)} J +\xi \Gamma^{++}_{(3)} J  \right) q^+_a\,.  \lb{Spin3qactionGauged}
\end{eqnarray}
The presence of constant {$\xi$} in the gauged Lagrangian  shows that off shell there are two possible interactions of {${\cal N}=2$} spin
${\bf 3}$ with the hypermultiplet. The coefficient {$\xi$} is a dimensionless coupling constant that measures the relative strength of these interactions. Recently we have checked that
on shell the {$\xi$} term does not contribute to the cubic vertex (at least in the bosonic sector), it survives only off shell and perhaps can play some
role in the quantum theory \cite{BIZ3}.\\

\noindent$\mathbf{Superspin\,\, 4}$. In the superspin 4 case the transformations of {$q^{+a}$} do not require including internal symmetry generators,
so they preserve {${\rm SU}(2)_{PG}$ }invariance.
The rigid symmetry
transformations are of the third order in derivatives and are well defined. Their localization on the analytic subspace admit 6 independent variations, but  only 3 of them
turn out to finally  matter
\begin{eqnarray}
&& \delta_1 q^{+a} = -\partial_{\alpha\dot{\alpha}} \partial_{\beta\dot{\beta}} \hat{\Lambda}^{\alpha\beta\dot{\alpha}\dot{\beta}} q^{+a}\,, \quad
\delta_2 q^{+a} =  -\hat{\Lambda}^{\alpha\beta\dot{\alpha}\dot{\beta}} \partial_{\alpha\dot{\alpha}} \partial_{\beta\dot{\beta}} q^{+a}\,,  \nonumber \\
&& \delta_3 q^{+a} = -\partial_{\alpha\dot{\alpha}} \partial_{\beta\dot{\beta}} \Omega^{\alpha\beta\dot{\alpha}\dot{\beta}} q^{+a}\,, \lb{Spin4Gauge}
\end{eqnarray}
where
\begin{eqnarray}
\hat{\Lambda}^{\alpha\beta\dot{\alpha}\dot{\beta}} = \lambda^{(\alpha\beta)(\dot{\alpha}\dot{\beta})M}\partial_M\,, \;
\Omega^{\alpha\beta\dot{\alpha}\dot{\beta}} = (-1)^{P(M)}(\partial_M\lambda^{(\alpha\beta)(\dot{\alpha}\dot{\beta})M}) \nonumber
\end{eqnarray}
and derivatives freely act to the right.

One can calculate
\begin{equation}
    \left(\delta_1 + \delta_2  + \delta_3\right) \mathcal{L}^{+4}_{free}
    =
     \frac{1}{2} q^{+a} [\mathcal{D}^{++}, \hat{\Lambda}_{(4)}] q^+_a. \nonumber
\end{equation}
This combination of variations is uniquely  distinguished by the requirement that it can be canceled by the gauge transformations of the spin ${\bf s}=4$ gauge multiplet.
The modified {$q^{+a}$} Lagrangian reads
\begin{equation}
        \mathcal{L}^{+4(s=4)}_{gauge} = -   \frac12\, q^{+a}  \left(\mathcal{D}^{++} + \hat{\mathcal{H}}^{++}_{(4)} \right) q^+_a\,,\lb{Spin4coupling}
\end{equation}
where
\begin{eqnarray}
&& \hat{\mathcal{H}}^{++}_{(4)} = h^{++(\alpha\beta)(\dot{\alpha}\dot{\beta})M}\partial_M\partial_{\alpha\dot{\alpha}}\partial_{\beta\dot{\beta}}\,, \nonumber \\
&& \delta \hat{\mathcal{H}}^{++}_{(4)} = [{\cal D}^{++},\hat{\Lambda}_{(4)}]\,,\quad \hat{\Lambda}_{(4)}=
\lambda^{(\alpha\beta)(\dot{\alpha}\dot{\beta})M}\partial_M\partial_{\alpha\dot{\alpha}}\partial_{\beta\dot{\beta}}\,.\label{Spin4Inv}
\eea

The hypermultiplet couplings of the {${\cal N}=2$} gauge multiplets of higher superspins ${\bf s}$ can be constructed quite analogously and they have the uniform
cubic structure, with gauge superfields and gauge parameters being the appropriate differential
operators of rank $({\bf s -1})$.

The whole consideration can be straightforwardly extended to several hypermultiplets.
The free Lagrangian of {$n$} hypermultiplets is written in the manifestly {${\rm USp}(2n)$} invariant form as
\bea
{\cal L}^{+4}_{free, n} = -\frac12 q^{+ A}{\cal D}^{++}q^+_A\,, \quad \widetilde{q^+_A} = \Omega^{AB} q^+_B\,, \;\; A = 1, 2, \ldots , 2n\,,\nonumber
\eea
where {$\Omega^{AB} = -\Omega^{BA}$} is {${\rm USp}(2n)$} invariant constant {$2n \times 2n$} symplectic metric. Its equivalent
complex form is given by
\bea
{\cal L}^{+4}_{free, n} \sim \tilde{q}^{+a}{\cal D}^{++}q^+_a -{\cal D}^{++} \tilde{q}^{+a} q^+_a\,, \quad  a= 1, 2, \ldots , n\,, \;\;q^+_A = (q^+_a, - \tilde{q}^{+ a})\, \nonumber
\eea
so that the manifest symmetry is reduced to {${\rm U}(n)= {\rm SU}(n)\times {\rm U}(1) \subset {\rm USp}(2n)$}, with respect to which {$q^+_a$} and {$\tilde{q}^{+ a}$}
transform in the fundamental and co-fundamental representations. One can identify the {${\rm U}(1)$} generator needed for the description of odd spins as
\bea
J q^+_a  = i q^+_a\,, \quad J \tilde{q}^{+ a} =  -i \tilde{q}^{+ a}\,, \nonumber
\eea
indicating that the original {${\rm USp}(2n)$} symmetry gets broken to {${\rm U}(n)$}. Some other options, with {${\rm SU}(n)$} being also broken,
are equally possible.

It should be pointed out that for all superspins, excepting ${\bf s= 1}$ and ${\bf s= 2}$, the cubic vertex is gauge invariant only
to the lowest order in gauge superfields. In general we cannot repeat the trick which works  in the superspin ${\bf 2}$ case: to replace the flat harmonic derivative
${\cal D}^{++}$ in the gauge transformation law of $\hat{\cal H}^{++}_{(s)}$ by the ``gauge-covariant'' one ${\cal D}^{++} + \hat{\cal H}^{++}_{(s)}$,
by analogy with \eqref{FullInv}. Because of presence of higher-degrees of derivatives in the differential operators $ \hat{\cal H}^{++}_{(s)}$ for $s\geq 3$,
such a naive replacement does not ensure full covariance.  Perhaps, for solving this problem one needs, from the very beginning, to add some kinds
of higher spin gauge transformations, with parameters properly composed from $\hat{\Lambda}_{(s)}$ (e.g., involving additional derivatives on the latter) and
so to permit the presence of all higher-spin potentials $\hat{\cal H}^{++}_{(s')}, s' > s$.\\

\noindent{\bf $\omega$- representation}. One more related point is that the $q^{+a}$ description of
the hypermultiplet yields the first-order formalism for physical scalar fields, while the correct second-order Lagrangian is restored only after
elimination of some auxiliary vector field component in $q^{+a}$. Simultaneously, besides the original trilinear vertex, there  appears the quartic component vertex with
two gauge fields and two physical scalars\footnote{This can be easily observed in the simplest spin ${\bf 1}$ case.}. The manifestly ${\cal N}=2$ supersymmetric version of passing from
the first-order formalism to the second-order one amounts to the so called $\omega$-formalism, with the following equivalent representation for $q^{+ a}$,
\bea
q^{+}_ a = u^{+}_a \omega + u^{-}_a L^{++}\,, \quad L^{++} = u^{+ a}q^+_a\,, \; \omega = -u^{- a}q^+_a\,,\lb{Fromqtoomega}
\eea
where $ L^{++} = L^{++}(\zeta_A)\,, \;\omega = \omega(\zeta_A)$. After this variable change, the Lagrangian in \eqref{Freeq}, modulo total harmonic derivative,
takes the form
\bea
\mathcal{L}^{+4}_{free} =  L^{++}{\cal D}^{++}\omega + \frac12 (L^{++})^2 \,. \lb{Freeqomega}
\eea
Varying with respect to the auxiliary superfield $L^{++}$, we obtain $L^{++} = -{\cal D}^{++}\omega$ and, after substituting this back in \p{Freeqomega},
come to the $\omega$ form of the hypermultiplet action
\bea
\mathcal{L}^{+4}_{free} \; \Rightarrow \; -\frac12 ({\cal D}^{++}\omega)^2\,.
\eea

Performing the same trick in the Lagrangians involving trilinear interaction vertex, we can find the $\omega$-representation for the corresponding
gauge-covariantized $q^{+ a}$ actions. This procedure is rather straightforward for the even spin ${\bf s}$ vertices, while it is more sophisticated for odd
${\bf s}$, because of the presence of ${\rm SU}(2)_{PG}$ breaking operator $J$ in this case. Though the second option also admits rather simple passing to the $\omega$ representation,
albeit with some more complicated superfield variable change as compared to \eqref{Fromqtoomega} \footnote{This change of variables is given in the book \cite{HSS}.}, here we limit
our consideration to the even superspins and, specifically, to  ${\bf s} = 2, 4$ only.

Making the change \eqref{Fromqtoomega} in the lagrangian \eqref{Spin2Hyper}, we obtain, up to a total derivative,
\bea
&& \mathcal{L}^{+4(s=2)}_{gauge} = {\cal D}^{++}\omega L^{++} + \frac12 (L^{++})^2  + L^{++} \big[\hat{\cal H}^{++}_{(2)}+ \frac12 \Gamma^{++}_{(2)}\big] \omega, \lb{Spin2omega} \\
&& \Gamma^{++}_{(2)} := (-1)^{P(M)} \partial_M h^{++ M}. \nonumber
\eea
Varying with respect to $L^{++}$ and substituting the result back in \p{Spin2omega}, we obtain the $\omega$ version of the superspin ${\bf s}=2$ covariantized $q^{+a}$ Lagrangian
\bea
\mathcal{L}^{+4(s=2)}_{gauge} \; \Rightarrow \;  -\frac12 \big[\big({\cal D}^{++} + \hat{\cal H}^{++}_{(2)}+ \frac12 \Gamma^{++}_{(2)}\big)\omega\big]^2\,. \label{omegaSpin2}
\eea
One can check that
\bea
\delta_{full}\omega = (\delta_1 + \delta_2)\omega = -\big(\hat\Lambda_{(2)} + \frac12 \Omega_{(2)}\big)\omega\,,
\eea
and
\bea
\delta_{full} \Gamma^{++}_{(2)} = \big({\cal D}^{++} + \hat{\cal H}^{++}_{(2)}\big)\Omega_{(2)} - \hat\Lambda_{(2)} \Gamma^{++}_{(2)}\,. \lb{Gamma2Tran}
\eea
Then it is easy to check that
\bea
\delta_{full} \mathcal{L}^{+4(s=2)}_{gauge} = \frac12 \big( \hat{\Lambda}_{(2)} + \Omega_{(2)}\big)\mathcal{L}^{+4(s=2)}_{gauge}\,.
\eea
This is a total derivative: the result of integrating by parts with respect to $\partial_M$ in the first term precisely cancels the second term.

The $\omega$-representation for the superspin ${\bf 4}$ gauge covariant action \p{Spin4Inv} can be obtained quite analogously:
\bea
\mathcal{L}^{+4(s=4)}_{gauge} \; \Rightarrow \;-\frac12 \big[\big({\cal D}^{++} + \hat{\cal H}^{++}_{(4)}+ \frac12 \Gamma^{++}_{(4)}\big)\omega\big]^2,\lb{Spin4omega}
\eea
where
\bea
\Gamma^{++}_{(4)}:= (-1)^{P(M)} \partial_{\alpha\dot\alpha}\partial_{\beta\dot\beta}\partial_M h^{++(\alpha\beta)(\dot\alpha\dot\beta)M}
\eea
(recall that undotted and dotted indices in $h^{++(\alpha\beta)(\dot\alpha\dot\beta)M}$ are assumed to be symmetrized with those hidden in the multi-index $M$).
The $\omega$ Lagrangian for generic even ${\bf s}$  is constructed as a straightforward generalization of \p{Spin2omega} and \p{Spin4omega}. Though the ${\bf s}\geq 4$  actions,
like their cubic $q^{+ a}$ cousins, are gauge invariant only up to the first order in the gauge superfields, they could serve as a good point of departure for constructing
the hypothetical fully gauge-invariant hypermultiplet actions, as they automatically incorporate the double gauge-superfield vertex giving rise to the correct
second-order component actions for the physical bosonic fields.

\section{Equations of motion and supercurrents}
In a recent paper \cite{BIZ3} there was performed a thorough study of the superfield equations of motion corresponding to a sum of the linearized actions for higher-spin ${\cal N}=2$ gauge superfields
and the covariantized hypermultiplet action. Varying with respect to unconstrained gauge analytic potentials yields both the equations with sources and the precise structure of these
sources in terms of the hypermultiplet superfields. On the other hand, varying with respect to $q^{+ a}$ yields the covariantized equations for the hypermultiplet. Partially solving
the latter fixes the harmonic dependence of $q^{+ a}$ and expresses the latter in terms of physical fields. In this section we concentrate on the first type of equations. It will
be convenient to restore the hypermultiplet coupling constants through the substitution $(\hat{H}^{++}_{(s)}, \,\Gamma^{++}_{(s)}) \,\Rightarrow \, (\kappa_s\hat{H}^{++}_{(s)}, \, \kappa_s\Gamma^{++}_{(s)})$,
with $[\kappa_s] = - {\bf s} + 1$ in mass units, in eqs. \p{Spin1coupl}, \p{Spin2Hyper}, \p{FullInv}, \p{Spin3qactionGauged}, \p{Spin4coupling}, \p{Spin2omega}, \p{Gamma2Tran}, \p{Spin4omega} and their higher ${\bf s}$ analogs.

The first non-trivial example of current superfields is supplied by the spin $\mathbf{2}$ case. Varying with respect to the unconstrained analytic potentials of the
linearized $\mathcal{N}=2$ supergravity,
 $$h^{++\alpha\dot{\alpha}}\,,\quad h^{++5}\,,\quad h^{++\alpha+}\,,\quad h^{++\dot{\alpha}+}\,,$$
 in the sum of pure superspin {\bf 2} gauge action and the superspin {\bf 2} hypermultiplet coupling  leads to the source-modified equations of motion\footnote{We use the standard definitions,
 $$(D^+)^2 := D^{+ \alpha}D^+_{\alpha}\,,\; :(\bar{D}^+)^2 := \bar{D}^{+}_{\dot\alpha}\bar{D}^{+\dot\alpha}\,,\; (D^+)^4 := \frac{1}{16}(D^+)^2(\bar{D}^+)^2\,.$$}:
\begin{eqnarray}
        && (D^+)^4 G^{--\alpha \dot{\alpha} }=  \frac{1}{2}  \kappa_2 J^{++\alpha \dot{\alpha}}\,, \nonumber
        \\
        &&(D^+)^4 G^{--5 } = \frac{1}{2}  \kappa_2 J^{++}\,,\nonumber
        \\
        &&(\bar{D}^+)^2 D^{+\alpha} G^{--}_{\alpha \dot{\alpha}}
        -
        (D^+)^2 \bar{D}^+_{\dot{\alpha}} G^{--5}
        = 2i \kappa_2  \mathcal{J}^+_{ \dot{\alpha}}\,, \nonumber
        \\
        &&(D^+)^2  \bar{D}^{+\dot{\alpha}} G^{--}_{\alpha \dot{\alpha}}
        +
        (\bar{D}^+)^2 D^+_{\alpha} G^{--5}
        =
        2i \kappa_2 \mathcal{J}^+_{\alpha }\,, \lb{EOM-spin 2}
\end{eqnarray}
where
\begin{equation}
({\rm a})\;    \mathcal{J}^+_{\alpha} = J^+_\alpha - 2i \bar{\theta}^{-\dot{\mu}} J^{++}_{\alpha\dot{\mu}}
    +
    2i \theta^-_\alpha J^{++}, \quad ({\rm b})\; \mathcal{J}^+_{\dot{\alpha}}
    =
    J^+_{\dot{\alpha}}
    +
    2i \theta^{-\mu} J^{++}_{\mu\dot{\alpha}}
    +
    2i \bar{\theta}^-_{\dot{\alpha}} J^{++}.
\end{equation}
Here
\begin{equation}
\begin{split}
     &J^{++}_{\alpha\dot{\alpha}} = - \frac{1}{2}
     q^{+a}  \partial_{\alpha\dot{\alpha}}  q^+_a\,,
     \quad
     J^{++} =  -
     \frac{1}{2} q^{+a} \partial_5 q^+_a \,,
     \\
     &J^+_\alpha = -\frac{1}{2} q^{+a} \partial^-_{\alpha} q^+_a\,, \quad
     J^+_{\dot{\alpha}} = -\frac{1}{2} q^{+a} \partial^-_{\dot{\alpha}} q^+_a\,.
     \end{split}
\end{equation}

Note that  all these  current superfields can be obtained  as proper projections of the single non-analytic ``master''  current superfield
$\mathcal{J}$ introduced (in the superconformal case) in ref. \cite{Kuzenko:1999pi}
\begin{equation}
    \begin{split}
    &\mathcal{J} := - \frac{1}{2} q^{+a} \mathcal{D}^{--} q^+_a,
    \qquad
    \mathcal{D}^{++} \mathcal{J} = 0.
    \end{split}
\end{equation}
Indeed, using the linearized equation of motion and the analyticity of $q^{+ a}$, we obtain
\begin{equation}
   \mathcal{J}^+_\alpha =  D^+_\alpha \mathcal{J}\,,
    \qquad
    \mathcal{J}^+_{\dot{\alpha}} = -\bar{D}^+_{\dot{\alpha}} \mathcal{J}\,,
\end{equation}
\begin{equation}
   J^{++}_{\alpha\dot{\alpha}} = -\frac{i}{2} D^{+}_\alpha \bar{D}^+_{\dot{\alpha}} \mathcal{J}\,, \qquad
    J^{++} = \frac{i}{4} (D^+)^2 \mathcal{J}
    =
      - \frac{i}{4} (\bar D^+)^2 \mathcal{J}\,.
\end{equation}
The geometric meaning of $\mathcal{J}$ becomes clear in the superconformal case, where it appears as a hypermultiplet current
superfield associated with the additional analytic gauge potential $H^{+4}$ \cite{Galperin:1987ek}. From this representation it immediately follows the common
conservation law
\be
{\cal D}^{++}J^{++}_{\alpha\dot{\alpha}} = {\cal D}^{++}\mathcal{J}^+_\alpha = {\cal D}^{++} J^{++} = 0\,,
\ee
which is compatible with the action of $D^{++}$ on the l.h.s. of eqs. \eqref{EOM-spin 2}, which yields zero as a consequence of the harmonic analyticity
and the zero-curvature condition \eqref{Spin2Zero}. The most important object is the supercurrent $J^{++}_{\alpha\dot{\alpha}}$ since it involves the energy-momentum tensor, fermionic spin-vector current
and $R$-symmetry vector current. It satisfies the additional conservation law
\be
\partial_{\alpha\dot{\alpha}}J^{++ \alpha\dot{\alpha}} = 0\,.
\ee

The generalization of \eqref{EOM-spin 2} to the case of ${\cal N}=2$ spin ${\bf s} =3$ is as follows
\begin{eqnarray}\label{EOM with currents spin 3}
        (D^+)^4 G^{--(\alpha\beta) (\dot{\alpha}\dot{\beta}) }= -\frac{1}{2}  \kappa_3 J^{++(\alpha\beta) (\dot{\alpha}\dot{\beta}) }\,,
        \nonumber \\
        (D^+)^4 G^{--5\alpha \dot{\alpha} } = -\frac{1}{2}  \kappa_3 J^{++\alpha \dot{\alpha}}\,,\nonumber
        \\
        (\bar{D}^+)^2 D^{+\beta} G^{--}_{(\beta\alpha) (\dot{\alpha}\dot{\beta})}
        -
        (D^+)^2 \bar{D}^+_{(\dot{\alpha}} G^{--5}_{\alpha \dot{\beta})}
        = -2i \kappa_3 \mathcal{J}^+_{\alpha (\dot{\alpha}\dot{\beta})}\,, \nonumber
        \\
        (D^+)^2  \bar{D}^{+\dot{\beta}} G^{--}_{(\alpha\beta) (\dot{\beta}\dot{\alpha})}
        +
        (\bar{D}^+)^2 D^+_{(\alpha} G^{--5}_{\beta) \dot{\alpha}}
        =
        -2i \kappa_3 \mathcal{J}^+_{(\alpha\beta) \dot{\alpha}}\,.
\end{eqnarray}
Here
\begin{subequations}\label{s=3 super current}
\begin{equation}
    J^{++}_{(\alpha\beta)(\dot{\alpha}\dot{\beta})}
    =
    - \frac{1}{2}
    q^{+a}  \partial_{(\alpha \dot{\alpha}} \partial_{\beta) \dot{\beta}} J q^+_a
    -
    \frac{1}{2} \xi  \,\partial_{(\alpha \dot{\alpha}} \partial_{\beta) \dot{\beta}} \left(q^{+a}  J  q^+_a \right),
\end{equation}
\begin{equation}
      J^{++}_{\alpha\dot{\alpha}}
    =
    -
    \frac{1}{2}
    q^{+a}\partial_{\alpha\dot{\alpha}} J \partial_5 q^+_a,
\end{equation}
\begin{equation}
    J^+_{(\alpha\beta)\dot{\alpha}} = -
    \frac{1}{2} q^{+a}\partial_{(\alpha\dot{\alpha}} \partial^-_{\beta)} J q^+_a - \frac{1}{2} \xi \partial_{(\alpha\dot{\alpha}} \partial^-_{\beta)} \left( q^{+a} J q^+_a \right),
\end{equation}
\begin{equation}
    J^+_{(\dot{\alpha}\dot{\beta})\alpha} = -
    \frac{1}{2} q^{+a}\partial_{\alpha(\dot{\alpha}} \partial^-_{\dot{\beta})} J q^+_a - \frac{1}{2} \xi \partial_{\alpha(\dot{\alpha}} \partial^-_{\dot{\beta})} \left( q^{+a} J q^+_a \right).
\end{equation}
\end{subequations}

Like in the ${\bf s}=2$ case, these current superfields can be related to a single ``master'' current superfield, this time the vector one,
\begin{equation}\label{s=3 precurrent}
    \mathcal{J}_{\alpha\dot{\alpha}} :=
    - \frac{1}{2} q^{+a} \mathcal{D}^{--} \partial_{\alpha\dot{\alpha}} J q^+_a
    - \frac{1}{2} \xi \,\mathcal{D}^{--} \partial_{\alpha\dot{\alpha}} \left( q^{+a}  J q^+_a \right),
\end{equation}
obeying, on the hypermultiplet mass shell,  the conservation law:
\begin{equation}\label{s=3 conservation law}
        \mathcal{D}^{++}    \mathcal{J}_{\alpha\dot{\alpha}} =  -\frac{1}{2} q^{+a} \partial_{\alpha\dot{\alpha}} J q^+_a - \xi \,
        \partial_{\alpha\dot{\alpha}}\left( q^{+a}  J q^+_a \right).
\end{equation}
We find
\begin{equation}
   \mathcal{J}^+_{(\alpha\beta)\dot{\alpha}} =  D^+_{(\alpha} \mathcal{J}_{\beta)\dot{\alpha}}\, ,
    \qquad
    \mathcal{J}^+_{(\dot{\alpha}\dot{\beta})\alpha} = -\bar{D}^+_{(\dot{\alpha}} \mathcal{J}_{\dot{\beta})\alpha}\,,
\end{equation}
\begin{equation}
    J^{++}_{(\alpha\beta)(\dot{\alpha}\dot{\beta})} = -\frac{i}{2} D^{+}_{(\alpha} \bar{D}^+_{(\dot{\alpha}} \mathcal{J}_{\beta)\dot{\beta})}\,,
\qquad J_{\alpha\dot{\alpha}}^{++} = \frac{i}{4}(D^+)^2 \mathcal{J}_{\alpha\dot{\alpha}}
    =
    -\frac{i}{4}(\bar D^+)^2 \mathcal{J}_{\alpha\dot{\alpha}}\,,
\end{equation}
where
\begin{subequations}
    \begin{equation}
        \mathcal{J}^+_{(\dot{\alpha}\dot{\beta})\alpha }
        =
        J^+_{(\dot{\alpha}\dot{\beta})\alpha }
        +
        2i \theta^{-\mu} J^{++}_{(\alpha\mu)(\dot{\alpha}\dot{\beta})}
        +
        2i \bar{\theta}^-_{(\dot{\alpha}}J^{++}_{\alpha\dot{\beta})},
    \end{equation}
    \begin{equation}
        \mathcal{J}^+_{(\alpha\beta) \dot{\alpha}}
        =
        J^+_{(\alpha\beta) \dot{\alpha}}
        -
        2i \bar{\theta}^{-\dot{\mu}} J^{++}_{(\alpha\beta)(\dot{\alpha}\dot{\mu})}
        +
        2i \theta^-_{(\alpha} J^{++}_{\beta)\dot{\alpha}}.
    \end{equation}
\end{subequations}
These superfields  satisfy the standard harmonic conservation laws, $\mathcal{D}^{++}J^{++}_{\alpha\dot\alpha} = 0\,,$ etc.

The construction outlined above can be directly extended to the case of an arbitrary ${\cal N}=2$ spin ${\bf s}$.

It is worth noting that the supercurrents in supersymmetric  ${\cal N}=2$ higher spin theories  have already been discussed in ref. \cite{Kuzenko:2021pqm}
(see also \cite{BHK} and recent papers \cite{KPR,Kuzenko:2023vgf}), without use of the harmonic superspace approach. These  theories seemingly
 correspond to the choice of special gauge for the analytic potentials. It is important to point out that in the papers just cited the current superfields
were constructed within the on-shell description of the hypermultiplet in terms of the properly constrained conventional ${\cal N}=2$ (or ${\cal N}=1$) superfields.
The basic distinction of the approach worked out in \cite{BIZ1,BIZ2} and \cite{BIZ3} is that there the off-shell description of both higher spin gauge ${\cal N}=2$
multiplets and matter hypermultiplets in terms of unconstrained harmonic analytic superfields was used.

\section{Conclusions}

The theory of {${\cal N}= 2$} higher spins  {${\bf s}\geq 3$} opens a new promising area
of applications of the harmonic superspace approach which earlier turned out to be indispensable for describing the more conventional {${\cal N}= 2$}
theories with the maximal spins {$s\leq 2$}. Like in the latter theories,
the basic principle  underlying the new higher-spin theories is the harmonic Grassmann analyticity: all basic gauge potentials
are unconstrained analytic superfields involving an infinite number of degrees of freedom off shell, before fixing WZ-type gauges.
\vspace{0.2cm}

{\bf Under way}:
\vspace{0.2cm}

\begin{itemize}
\item The natural next steps are the construction and analysis of {$4D, \mathcal{N}=2$} higher-spin superconformal theory in the HSS approach
as a generalization of the HSS formulation of conformal ${\cal N}=2$ supergravity \cite{Galperin:1987ek}
and its couplings to the hypermultiplets. This study is in progress now.
\vspace{0.3cm}

\item It still remains to construct ${\cal N}=2$ higher spins  in AdS  background (and more general conformally flat backgrounds). This problem should
be intimately related to that of constructing conformal ${\cal N}=2$ higher-spin theory.
\vspace{0.3cm}

\item The ${\cal N}=2$ higher spin theory described above is a generalization, to higher spins, of the linearized minimal Einstein ${\cal N}=2$ supergravity
\cite{MinN2SG}. The HSS formulations of other versions of Einstein ${\cal N}=2$ supergravity (see \cite{HSS} and \cite{Ivanov}) should give rise to
some different versions of ${\cal N}=2$ higher spin theory. All these versions should be related to conformal ${\cal N}=2$ higher-spin theory through
the proper compensator mechanisms.
\vspace{0.3cm}

\item Until now, it is still unclear how to describe {${\cal N}=2$} supersymmetric half-integer spins.
\vspace{0.3cm}

\item How to pass from the linearized theory to its full nonlinear version? At present, the latter is known only
for {${\bf s}\leq 2$} ({${\cal N}=2$} super Yang - Mills and {${\cal N}=2$} supergravities). Solving this problem for {${\bf s}\geq 3$} will seemingly require  accounting
for ALL ${\cal N}=2$ higher  superspins simultaneously. One could expect to encounter new supergeometries on this road.
\vspace{0.3cm}

\item Since we are aware of the free Lagrangians and equations of motion for the unconstrained analytic higher-spin gauge ${\cal N}=2$ superfields, we can construct superfield
propagators for them  and study, e.g., the simplest quantum corrections to the free Lagrangians and hypermultiplet couplings. This should rely on the appropriate
background superfield method in harmonic superspace (generalizing that for the standard ${\cal N}=2$ gauge superfield-hypermultiplet systems, see, e.g., \cite{EffN2}).

\end{itemize}

\setcounter{equation}{0}

\noindent {\bf Acknowledgements.} I thank the Organizers of the A.A. Slavnov Memorial Conference for
inviting me to give this talk and kind hospitality in Steklov Institute in Moscow. I also thank my co-authors Ioseph Buchbinder and Nikita Zaigraev, on the joint papers with
whom this talk is essentially based. I acknowledge support from the Russian Science Foundation, project No 21-12-00129. It is also worth to thank the anonymous referee for useful
remarks aimed at making the exposition more complete.

\end{document}